\documentclass[aps,article,a4paper,10pt,notitlepage,footinbib,superscriptaddress,longbibliography]{revtex4-1}
\usepackage[english]{babel}
\usepackage{lmodern}
\usepackage[latin9]{inputenc}
\usepackage{amssymb,amsmath,amsfonts}
\usepackage{textcomp}
\usepackage{braket}
\usepackage{ulem}
\usepackage{soul}
\usepackage{graphicx}% Include figure files
\usepackage{dcolumn}% Align table columns on decimal point
\usepackage{bm}% bold math
\usepackage{xcolor}
\usepackage{soul}
\usepackage{hyperref}
\usepackage{comment}
\usepackage{float}
\usepackage{soul}

\definecolor{bostonuniversityred}{rgb}{0.8, 0.0, 0.0}
\definecolor{massimiliano}{RGB}{0,0,255}

\begin{document}

%tentative title
\title{Transverse self-propulsion enhances aggregation of active dumbbells}

\author{Pasquale Digregorio}
\affiliation{Dipartimento  Interateneo di  Fisica,  Universit\`a  degli  Studi  di  Bari \\ and INFN, Sezione  di  Bari, via  Amendola  173,  Bari,  I-70126,  Italy}

\author{Claudio Basilio Caporusso}
\affiliation{Dipartimento  Interateneo di  Fisica,  Universit\`a  degli  Studi  di  Bari \\ and INFN, Sezione  di  Bari, via  Amendola  173,  Bari,  I-70126,  Italy}

\author{Lucio Mauro Carenza}
\affiliation{Dipartimento  Interateneo di  Fisica,  Universit\`a  degli  Studi  di  Bari \\ and INFN, Sezione  di  Bari, via  Amendola  173,  Bari,  I-70126,  Italy}

\author{Giuseppe Gonnella}
\affiliation{Dipartimento  Interateneo di  Fisica,  Universit\`a  degli  Studi  di  Bari \\ and INFN, Sezione  di  Bari, via  Amendola  173,  Bari,  I-70126,  Italy}

\author{Daniela Moretti}
\affiliation{Dipartimento  Interateneo di  Fisica,  Universit\`a  degli  Studi  di  Bari \\ and INFN, Sezione  di  Bari, via  Amendola  173,  Bari,  I-70126,  Italy}

\author{Giuseppe Negro}
\affiliation{SUPA, School of Physics and Astronomy, University of Edinburgh EH9 3FD, UK}

\author{Massimiliano Semeraro}
\affiliation{Dipartimento  Interateneo di  Fisica,  Universit\`a  degli  Studi  di  Bari \\ and INFN, Sezione  di  Bari, via  Amendola  173,  Bari,  I-70126,  Italy}

\author{Antonio Suma}
\affiliation{Dipartimento  Interateneo di  Fisica,  Universit\`a  degli  Studi  di  Bari \\ and INFN, Sezione  di  Bari, via  Amendola  173,  Bari,  I-70126,  Italy}

\begin{abstract}
We investigate a two-dimensional system of active Brownian dumbbells using molecular dynamics simulations. In this model, each dumbbell is driven by an active force oriented perpendicular to the axis connecting its two constituent beads. We characterize the resulting phase behavior and find that, across all values of activity, the system undergoes a phase separation between dilute and dense phases. The dense phase exhibits hexatic order and, for large enough activity, we observe a marked increase in local polarization, with dumbbells predominantly oriented towards the interior of the clusters. Compared to the case of axially self-propelled dumbbells, we find that the binodal region is enlarged towards lower densities at all activities. This shift arises because dumbbells with transverse propulsion can more easily form stable cluster cores, serving as nucleation seeds, and show a highly suppressed escaping rate from the cluster boundary. Finally, we observe that clusters exhibit spontaneous rotation, with the modulus of the angular velocity scaling as $\omega\sim r_g^{-2}$, where $r_g$ is the cluster's radius of gyration. This contrasts with axially propelled dumbbells, where the scaling follows $\omega\sim r_g^{-1}$. We develop a simplified analytical model to rationalize this scaling behavior.
\end{abstract}

\maketitle

\section{Introduction}
\label{sec:intro}

Active matter describes a diverse range of systems composed of agents that autonomously convert energy into directed motion or mechanical stresses, thus continuously operating far from equilibrium~\cite{bechinger2016,Ramaswamy10,Vicsek12,Marchetti13,gonn15,Elgeti15,Care2019,te2025metareview}. Spanning across physics, biology, and materials science, active matter includes examples as varied as swimming microorganisms, cellular tissues, animal groups, and synthetic self-propelled particles. These active systems give rise to distinctive collective phenomena~\cite{sanchez2012,demagistris2015,cagnetta2017large,kumar2018,Doostmo2018,Santillan2018,negro2019hydrodynamics,carenza2020soft,giordano2021activity,Gompper_2020,Fodor23,Dauchot23,Caporusso-chiral23,Wiese23,Caprini23,Favuzzi03072021,Bayram23,Caprini2020}, such as clustering and spontaneous flow patterns, which cannot be understood within the classical framework of equilibrium statistical mechanics. 

From a theoretical perspective, particle-based models play a crucial role in understanding active systems. A prominent example is provided by active Brownian particles~\cite{Digregorio18,Digregorio19,Fily12,Redner13,negro2022hydrodynamic,Caporusso20,Caporusso22,semeraro2024}, where each particle experiences a self-propulsion force of fixed magnitude, whose orientation evolves stochastically due to rotational diffusion. Despite their apparent simplicity, these models can capture rich and intriguing nonequilibrium behaviors. For instance, purely repulsive active Brownian disks exhibit a diverse phase diagram and, beyond a critical propulsion strength, undergo motility-induced phase separation (MIPS)-a spontaneous aggregation driven solely by activity, rather than attractive interactions~\cite{Fily12,Redner13,Digregorio18}.

When considering asymmetric Brownian particles, the phenomenology becomes even richer, as their local interactions will depend significantly on the particles orientation and  will influence their collective behavior, introducing additional complexity such as directional alignment, shape-dependent interactions, and orientational order~\cite{Moran22,Peruani2006,Ginelli2010,Yang2010,Dijkstra19,Bar20,paoluzzi_glass}. In this context, active dumbbells-minimal models of rigid molecules composed of two connected particles-are relevant as a paradigmatic asymmetric particle model~\cite{Suma14,Cugliandolo2017,Petrelli18,Caporusso23,caporusso2024phase,solon2015pressure,Fily18,Pirhadi21,Joyeux16,Joyeux17,Suma14b,Suma14c,Tung16,Schwarz-Linek12,Siebert2017}. Similarly to active disks, active dumbbells display MIPS~\cite{Suma14,Cugliandolo2017,Petrelli18}. At the same time, many additional properties arise. For instance, dumbbells tend to form a single global hexatic and dense phase coexisting with the liquid, and this regime continuously extends to the passive dumbbell case, where a first-order liquid-hexatic transition is present~\cite{Cugliandolo2017,Petrelli18}. Furthermore, the dumbbells shape allow particles to better interlock with each other, phase-separating at smaller densities than active disks, and spontaneously forming clusters with polar order~\cite{Suma14,Petrelli18}; higher activities provide stronger polarizations. This arrangement results in clusters rotating with a peculiar dynamics~\cite{Suma14,Petrelli18,caporusso2024phase}.  

Much has been done to characterize active dumbbells either with axial self-propulsion or with a self-propulsion direction that diffuses independently from axis of the molecule~\cite{Siebert2017}. However, it has been shown both in experiments and numerical studies that configurations with transverse self-propulsion are relevant for a variety of potential applications, like target delivery or cargo transport~\cite{Fei2017,Lee2021,Parisi2018}. Thus, it is important to characterize general aggregation properties and cluster behavior if the active force is applied in a transverse direction (perpendicular to the dumbbell's head-to-tail direction). Here, we characterize the phase diagram of a two-dimensional system composed by dumbbells self-propelled along the transverse direction, through the use of molecular dynamics simulations. We show that, at all values of activity considered, the system undergoes a phase transition between dilute and dense phases, characterized through the local density distribution. The binodal region connects directly to the case of no activity, which was established in Ref.~\cite{Cugliandolo2017}. We find that the dense phase exhibits also hexatic ordering and an emergence of local polarization for sufficiently high activity. Due to the active torque present inside clusters, these are able to display a rotational motion, with the modulus of the angular velocity, $\omega$, scaling with the radius of gyration, $r_g$, as $\omega\sim r_g^{-2}$. We also discuss how these results compare to those obtained in the case of dumbbells propelled along their axis~\cite{Suma14,Cugliandolo2017,Petrelli18,caporusso2024phase}. First, we find that, for dumbbells propelled with a transverse force, the binodal region is expanded towards lower densities at all activities. Then, we develop a simplified model to highlight relevant differences in between the two models, concerning the scaling of $\omega$ with respect to $r_g$, is different.

%%%%%%%%%%%%%%%%%%%%%%%%%%%%%%%%%%%%%%%%%%
\section{Model and numerical methods}
\label{sec:model}

\begin{figure}[t!]
  \centering
  \includegraphics[width=\textwidth]{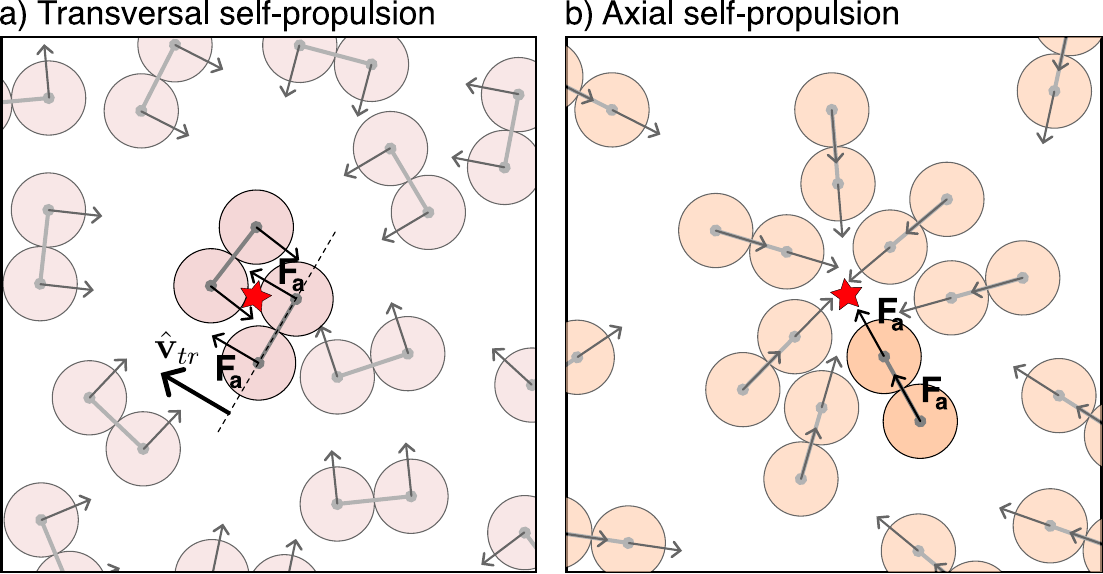}
  \caption{Schematic representation of dumbbells with (a) transverse self-propulsion and (b) axial self-propulsion. The active force vector acting on each bead is highlighted with an arrow. These representations depict nucleation events for MIPS, with multiple dumbbells colliding with their self-propulsion directed towards a common center, indicated by a red star. In case (a), a nucleation event can be composed by only two dumbbells, while in case (b) more dumbbells are needed.
  }
  \label{fig:trasv_dumbbell}
\end{figure}

We study a two-dimensional system composed of $N$ dumbbells. Each dumbbell is modeled as a diatomic molecule formed by two identical circular beads, rigidly linked, resulting in a total of $2N$ beads. Each bead has a diameter $\sigma_{\rm d}$ and mass $m_{\rm d}$. The distance between the centers of the two beads constituting a dumbbell is fixed to $\sigma_{\rm d}$. To each dumbbell, we assign a unit vector $\hat{\mathbf{v}}_{tr}$, orthogonal to the main axis connecting the dumbbell's head and tail. The active force is applied along this transverse direction (see Fig.~\ref{fig:trasv_dumbbell}(a) for a sketch).

The beads are labeled by $i=1,\dots,2N$, with positions denoted by ${\mathbf r}_i$, and their dynamics obey the Langevin equation:
\begin{equation}
  m_{\rm d} \ddot {\mathbf r}_i=-\gamma_{\rm d}\dot {\mathbf r}_i -{\boldsymbol \nabla}_i U+
  {\mathbf F}_{a}+
  \sqrt{2k_BT\gamma_{\rm d}} \, {\boldsymbol \eta}_i \; ,
  \label{eq:bd}
\end{equation}
where ${\boldsymbol \nabla}_i=\partial_{\mathbf{r}_i}$, $\gamma_{\rm d}$ and $T$ are the friction coefficient and the temperature of the thermal bath, respectively, and $k_B$ is the Boltzmann constant. ${\boldsymbol \eta}_i$ is a Gaussian white noise with $\langle\eta_{ia}(t)\rangle=0$, and $ \langle\eta_{ia}(t_1)\eta_{jb}(t_2)\rangle =\delta_{ij}\delta_{ab}\delta(t_1-t_2)$, with $a,b=1,2$ the label for the two spatial coordinates. Note that the thermal noise affects both the translational and rotational degrees of freedom of the dumbbells~\cite{Suma14b,Suma14c}.

Excluded volume interactions are mediated by the Mie potential~\cite{Mie1903}:
\begin{equation}
    U_{\rm Mie}(r) = \left\{ 4\epsilon \left[ \left(\frac{\sigma}{r} \right)^{2n} - \left(\frac{\sigma}{r}\right)^{n} \right] +\epsilon \right\} \Theta(2^{1/n}\sigma-r),
\end{equation}
with $\Theta$ the Heaviside function $\Theta(0)=0$), $\sigma$ and $\epsilon$ the length and energy scales of the potential. The potential is truncated at its minimum ($r=2^{1/n}\sigma$) so that it is purely repulsive. $n=32$ such that the potential is very steep, in order to be close to the hard-disk limit~\cite{Digregorio18}. Moreover, we choose $2^{1/n}\sigma=\sigma_{\rm d}$ so that the minimum of the potential equals the bead diameter. The rigid bond between each dumbbell's bead is enforced via the RATTLE numerical integration scheme~\cite{rattle}. The active force ${\mathbf F}_{a}$ has constant modulus $F_{a}$ and direction given by $\hat{\mathbf{v}}_{tr}$, computed for each dumbbell, and changing in direction through rotational diffusion undergone by the dumbbell.

The system evolves in a square box of side $L$, so that the surface fraction covered by the beads is
\begin{equation}
  \phi=\frac{N\pi\sigma_{\rm d}^2}{2L^2}
  \; ,
\end{equation}
The P\'eclet number,
\begin{equation}
  \mbox{Pe} = \frac{2\sigma_{\rm d} F_{a}}{k_BT}
  \; ,
\end{equation}
represents the dimensionless ratio between the work done by the active force and the thermal energy scale $k_BT$ and will be used as a control parameter for the system's activity.

We consider systems with $N=256^2/2$ dumbbells, and we span $\phi \in [0.1,0.9]$, setting $L$ in order to have the target density at given $N$. We use as system's units the mass $m_{\rm d}$, the diameter $\sigma_{\rm d}$ and the typical potential energy $\epsilon$~\cite{allen}, which are set to one, along with $k_B$. In these units, the molecular dynamics time units is $\tau_{MD}=\sqrt{m_{\rm d}\sigma_{\rm d}^2/\epsilon}$. We fix $\gamma_{\rm d}=10$ and $k_BT=0.05$. We also consider Pe $= 1$ to $100$, controlled by the value of $F_a$. The system is evolved through the Verlet algorithm using the LAMMPS software package~\cite{LAMMPS}. The simulation timestep is $10^{-3} \tau_{MD}$, with typical simulations of duration $3 \times 10^8\tau_{MD}$.

%%%%%%%%%%%%%%%%%%%%%%%%%%%%%%%%%%%%%%%%%%
\section{Results}

%%%%%%%%%%%%%%%%%%%%%%%%%%%%%%%%%%%%%%%%%%
\subsection{Phase diagram}

\begin{figure}[t!]
\begin{center}
    \includegraphics[width=\textwidth]{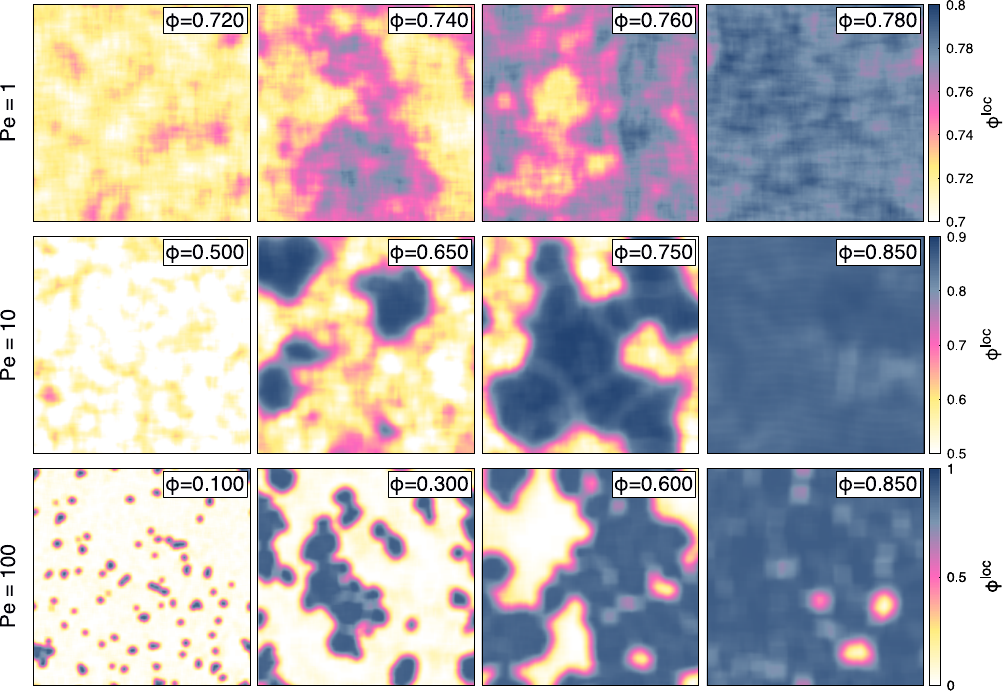}
    \caption{Snapshots of transverse self-propelled dumbbells, colored according to the local surface fraction $\phi^{loc}$ of transverse self-propelled dumbbells, defined in the text. Snapshots are shown for Pe $=1,10,100$ (first, middle and last rows, respectively) and global $\phi$ values displayed in each snapshot. The range of $\phi$ spans the region where a coexistence occurs between a dense and a dilute phase. The range of values of $\phi^{loc}$ is the same for all the snapshots in each row, and is displayed in the color bar on the right side.
    }
    \label{fig:phase_diagram_confs}
\end{center}
\end{figure}

Fig.~\ref{fig:phase_diagram_confs} shows a series of snapshots of system evolved at Pe $= 1,10,100$, with colors representing the local density $\phi^{loc}$, computed as a local average of the surface fraction over a grid of size $2 \sigma_d$, averaging over a coarse-graining length of $20 \sigma_d$.  The snapshots show the range in density where the system phase-separates into high and low density regions.  This range varies with the value of activity, but at the same time is observed at all Pe values considered. In particular, the range of densities is observed to increase with Pe. At Pe $=1$, we observe the transition to occur at high densities, between $\phi=0.725$ and $\phi=0.765$; at Pe $=10$ the range increases, from $\phi=0.600$ to $\phi=0.870$. At larger activity, corresponding to Pe = 100, we observe phase separation between $\phi=0.05$ and $\phi=0.890$. 

A special limiting case is the one of Pe $=0$, or no activity. This corresponds to a passive system of rigid dumbbells, which has already been characterized extensively in Ref.~\cite{Cugliandolo2017}. In this limit, it has already been established that the system undergoes a phase-separation between a low and a high density phase in the interval $\phi \in [0.730:0.756]$~\cite{Cugliandolo2017}.

\begin{figure}[t!]
  \centering
  \includegraphics[width=\textwidth]{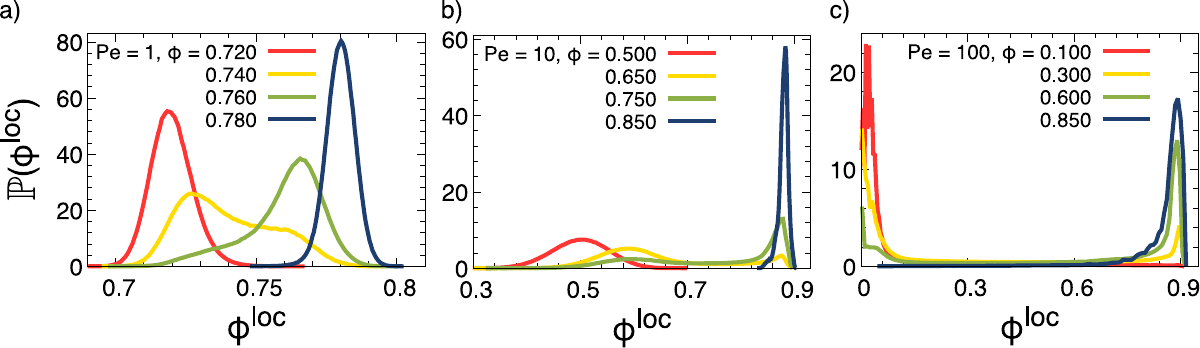}
  \caption{Probability distribution of the local surface fraction $\phi^{\rm{loc}}$, averaged over independent configurations, across the coexistence region for (a) Pe $=1$, (b) $10$ and (c) $100$. The values of global surface fraction are displayed in the key.
  }
  \label{fig:sf_pdfs}
\end{figure}

To corroborate these observations, we measured the probability distribution of the local density $\mathbb{P}(\phi^{loc})$. Fig.~\ref{fig:sf_pdfs} presents such distributions for the cases of Fig.~\ref{fig:phase_diagram_confs}. For Pe $=1$, we observe a single-peaked density distribution for $\phi \le 0.725$ and $\phi \ge 0.765$, the red and blue curves in Fig.~\ref{fig:sf_pdfs} (a), respectively. For all the cases with global density in between these two limits, the distribution is instead bimodal, with the two peaks remaining at fixed values of $\phi_{low}=0.725$ and $\phi_{high}=0.765$, which correspond to the values of the two coexisting densities, while the height of the two peaks varies as we move within these two limits. A similar situation occurs also at all the other Pe considered, see Fig.~\ref{fig:sf_pdfs} (b,c). 

Finally, by measuring the values of the coexisting densities, one can establish the binodal region in the phase space $\phi-$Pe where the phase-separation occurs. These values are plotted in Fig.~\ref{fig:phase_diagram}, where the low and high limits are shown with black empty and filled square symbols, respectively, alongside the values at Pe $=0$. We find here that the coexistence region, starting from Pe $=0$, opens up upon increasing activity. In addition, we plot a red and blue line, representing points where the surface occupied by the dense phase is 25\% and 50\% of the total system's area, respectively.

%%%%%%%%%%%%%%%%%%%%%%%%%%%%%%%%%%%%%%%%%%
\subsection{Comparison between transverse and axially-driven dumbbells and aggregation mechanism}

\begin{figure}[t!]
  \centering
  \includegraphics[width=\textwidth]{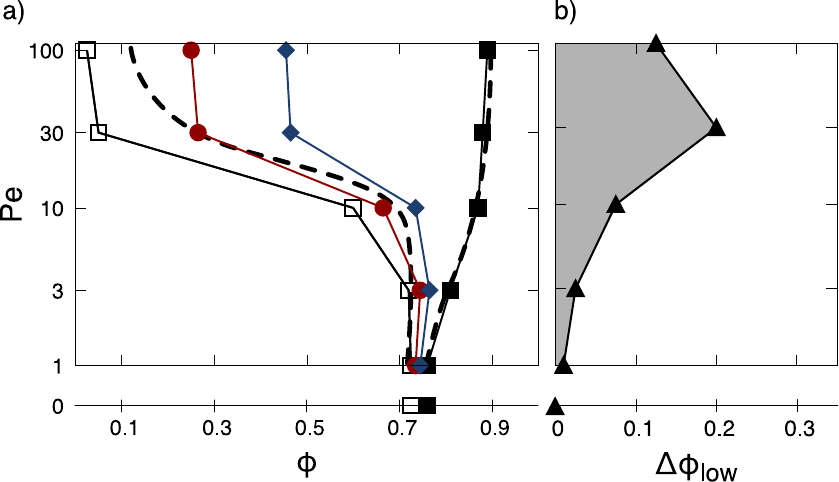}
  \caption{(a) The $\phi-$Pe phase diagram of the dumbbell system with transverse propulsion. The black empty and filled square symbols represent the values of the two coexisting densities, $\phi_{low}$ and $\phi_{high}$, respectively. The region in between these points is the binodal region, enclosing the portion of the phase diagram where we observe liquid-hexatic phase separation. The red and blue symbols indicate points on the curve where the surface occupied by the dense phase is 25\% and 50\% of the total one, respectively. The black dashed line indicates the two coexisting densities for axially self-propelled dumbbells, as established in~\cite{Cugliandolo2017}, for comparison. (b) Difference $\Delta\phi_{low}$ between the values of the low density branch $\phi_{low}$ computed for  axially-propelled dumbbells and those computed for dumbbells with transverse propulsion.
  }
  \label{fig:phase_diagram}
\end{figure}

We now compare the  behavior of active dumbbells driven by a transverse active force, with those driven by an axial active force, Fig.~\ref{fig:trasv_dumbbell}. In the latter system, it was found a similar phase separation behavior~\cite{Cugliandolo2017,Petrelli18}, with specific ranges of densities where a high and a low density regions coexist, again with the coexisting region starting directly at Pe $=0$, and continuously opening-up upon increasing Pe. The limit of the binodal region, reproduced in Fig.~\ref{fig:phase_diagram} as a dashed line, are similar in shape as those of dumbbells with a transverse active force. At the same time, however, the curve of the limiting low density is shifted towards higher values in axially-driven dumbbells. This shift increases with increasing Pe, as shown in Fig.~\ref{fig:phase_diagram} (b), which reports the difference $\Delta \phi_{low}$ between the values of the low density branch $\phi_{low}$ computed for axially-propelled dumbbells and those computed for dumbbells with transverse propulsion.
This quantity in facts peaks around Pe $=30$, then shrinking back for even higher activity.

It is worth noting that the widening of the binodal region at higher Pe for dumbbells with transverse propulsion is in contrast with the results of Ref.~\cite{Siebert2017}, where instead dumbbells have an active force on each bead, acting in the same direction, but diffusing independently from the dumbbell axis. This observation highlights the relevance of constraining the active force to the particle shape, regarding the aggregation properties. As a matter of fact, in all types of dumbbells considered, when activity becomes relevant with respect to noise, Pe$\gg1$, the cluster nucleation occurs in the same way, with several dumbbells being propelled towards a common center, colliding together and getting stuck for long enough so that other dumbbells arrive and trap them (this is the typical picture to explain the motility induced phase separation). 

However, different propulsion directions lead to different scenarios for two different reasons. The first lies in how the particles hit each other. Axial dumbbells forming a nucleation seed should have their axis pointing towards a common center, Fig.\ref{fig:trasv_dumbbell}. Thus, more than two dumbbells are needed to form a nucleus. 
For transverse self-propulsion, if two dumbbells collide along their transverse direction with self-propulsion oriented towards each other, they naturally arrange in a quadruplet instead, which then becomes the nucleation point. For dumbbells with a randomly changing propulsion direction it should be a mixed situation instead. 

The second reason is how the dumbbells can disengage the cluster. The idea is that particles at the cluster's surface should be able to orient the active force away from the cluster, in order to escape~\cite{Redner13}. This can occur either from collisions with other particles, or by the active force changing orientation through diffusion. In dumbbells with self-propulsion orientation unrelated to the dumbbell axis, this diffusion occurs independently of the dumbbells arrangement in the cluster, and thus the behavior becomes more similar to that of active colloids, as particles are more prone to escape and thus one expects a smaller binodal region~\cite{Siebert2017}. Instead, when the active force is locked in the axial or transverse direction, dumbbells detachment from clusters becomes much more difficult, as beads are interlocked together inside clusters, hence a larger binodal region. Moreover, a dumbbell with a transverse active force touches the cluster surface through a larger area, and thus is more stably interlocked to others beads, thus the additional enhancement of the binodal region. 
This mechanism is especially important at intermediate Pe, Fig.~\ref{fig:phase_diagram} (b).

%%%%%%%%%%%%%%%%%%%%%%%%%%%%%%%%%%%%%%%%%%
\subsection{Hexatic ordering of dense phase}

\begin{figure}[t!]
\begin{center}
    \includegraphics[width=\textwidth]{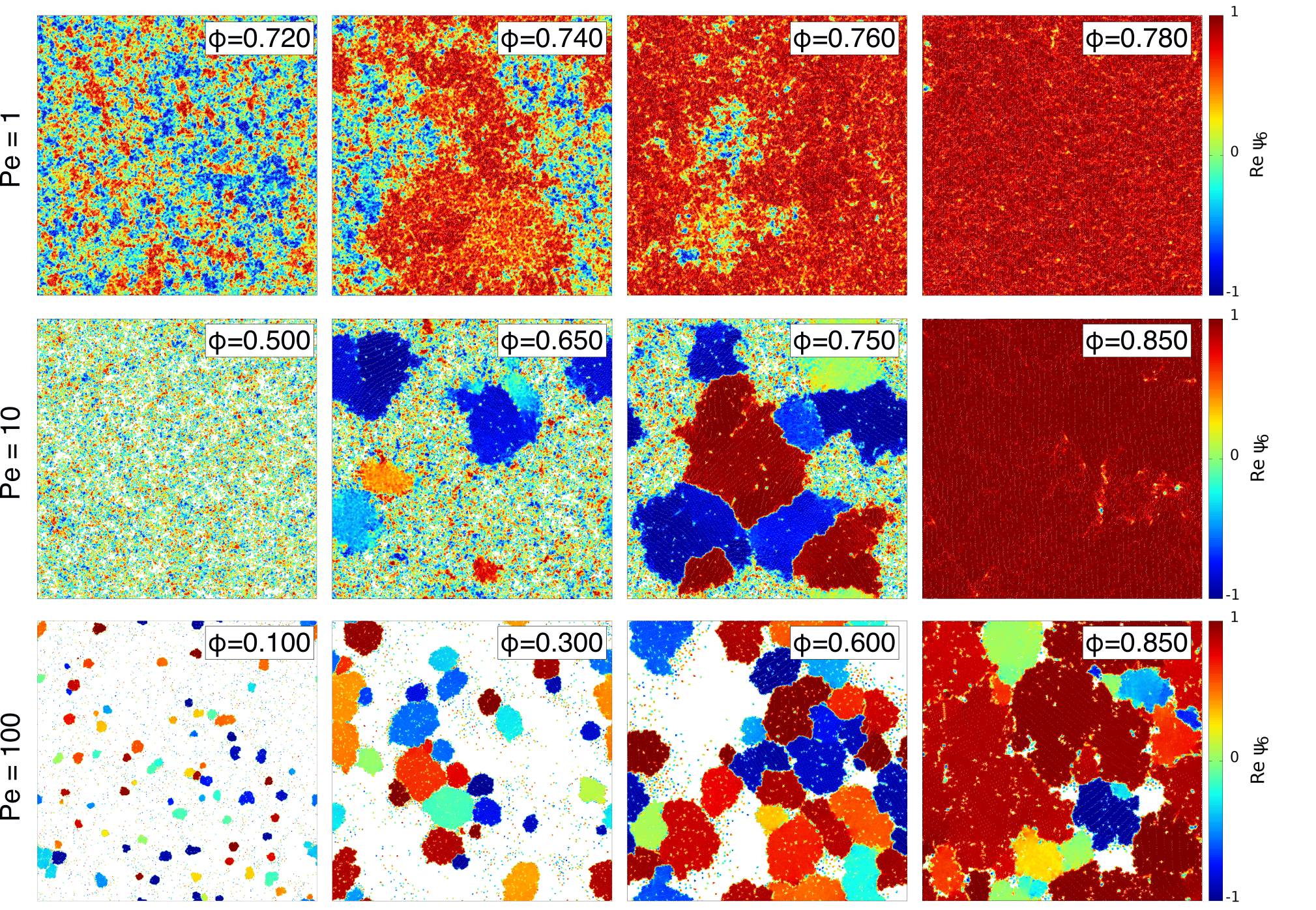}
    \caption{Snapshots for Pe $=1,10,100$ (first, middle and last rows, respectively) and global $\phi$ values displayed in each snapshot (same values as Fig.~\ref{fig:phase_diagram_confs}). Particles are colored according to real component of the local hexatic order parameter $\psi_{6,j}$  (Eq.~\ref{eq:hexatic}). The color code is reported on the right side. 
    }
    \label{fig:hexatic_confs}
\end{center}
\end{figure}

An important characteristic of the dense phase, also present in axially-propelled dumbbells\cite{Cugliandolo2017,Petrelli18}, is hexatic ordering. To characterize this order, we computed the local hexatic order parameter $\psi_{6,j}$, defined for the generic bead $j$ as: 
\begin{equation}
\label{eq:hexatic}
 \psi_{6,j} \equiv \frac{1}{N_j}\sum_{k=1}^{N_j} e^{i6\theta_{jk}},
\end{equation}
where the sum runs over the $N_j$ first neighbor beads and $\theta_{jk}$ is the orientation of the vector $\mathbf{r}_{jk}=\mathbf{r}_k-\mathbf{r}_j$ joining the two beads, with respect to the $x-$axis.

In Fig.~\ref{fig:hexatic_confs}, we show snapshots of system, for the same cases as in Fig.~\ref{fig:phase_diagram_confs}, with particles colored according to the real part of the local hexatic parameter. At Pe $=1$, the dense phase in the coexisting region, as well as the system at $\phi>0.765$, displays a clear hexatic order. In the representation used, uniformly colored regions correspond to hexatically ordered domains, i.e. regions in which beads tend to locally arrange according to a hexagonal lattice. At larger Pe $=10$, the global hexatic ordering is also clear looking at $\phi=0.85$, a density above the coexistence line, and similarly for Pe$=100$ (not shown). At the same time, in the coexistence region at both Pe=$10, 100$ the dense phase is composed by interlocked clusters, each hexatically ordered, but with different hexatic orientations. This effect appears also in Ref.~\cite{Cugliandolo2017}, where it was attributed to a metastable arrangement of clusters, expecting that for sufficiently long times  the dense phase will eventually have a single hexatic order. 

\begin{figure}[t!]
\begin{center}
    \includegraphics[width=\textwidth]{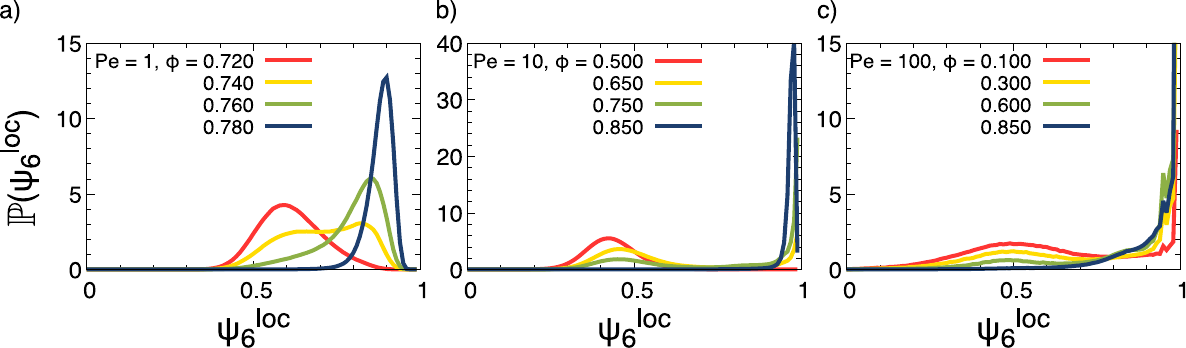}
    \caption{Probability distribution of the local hexatic parameter $\psi_6^{\rm{loc}}$, computed as the local average of the hexatic parameter (Eq.~\ref{eq:hexatic}), over a square grid of size $2 \sigma_d$, with coarse-graining length $5 \sigma_d$, averaged over independent configurations, across the coexistence region for Pe $=1, 10, 100$, respectively the left, middle, and right panels. Density values are displayed in the key.
    }
    \label{fig:hexatic_pdfs}
\end{center}
\end{figure}

A more quantitative way to characterize the hexatic order is to look at the probability distribution of the modulus of the local hexatic parameter, $\mathbb{P}(\psi_6^{loc})$, averaged over a grid of size $5 \sigma_d$., Fig.~\ref{fig:hexatic_pdfs}. The distributions have a similar shape as the local density ones of Fig.~\ref{fig:sf_pdfs}: below the coexistence region we observe a single-peaked density distribution; in between the coexisting region the distribution is instead bimodal, with the high-value peak pointing towards values near $1$, corresponding to the hexatic ordered region. At densities above the coexistence line we obtain again a single-peaked distribution, again at values near $1$.

%%%%%%%%%%%%%%%%%%%%%%%%%%%%%%%%%%%%%%%%%%
\subsection{Polarization of dense phase}

We now turn to the characterization of the internal organization of the dense phase. In axially-propelled dumbbells, the tendency to compact into hexatically ordered clusters typically implies that the dumbbells arrange along reciprocal angles multiples of $\pi/3$. At high Pe, this tendency is matched by an increase in the internal cluster polarization, meaning that dumbbells tend to be more aligned with each other, forming spiral patterns~\cite{Suma14, Petrelli18, caporusso2024phase}. At the same time, highly compacted dumbbells display translational and rotational persistent motion, that can be attributed to the total active force on the cluster center of mass and of total active torque~\cite{caporusso2024phase}. Here, given that transverse-propelled dumbbells follow again a hexatic arrangement and are even more prone to aggregate, Fig.~\ref{fig:trasv_dumbbell}, we expect that clusters in the dense phase display similar mechanical properties. 

\begin{figure}[t!]
  \centering
  \includegraphics[width=\textwidth]{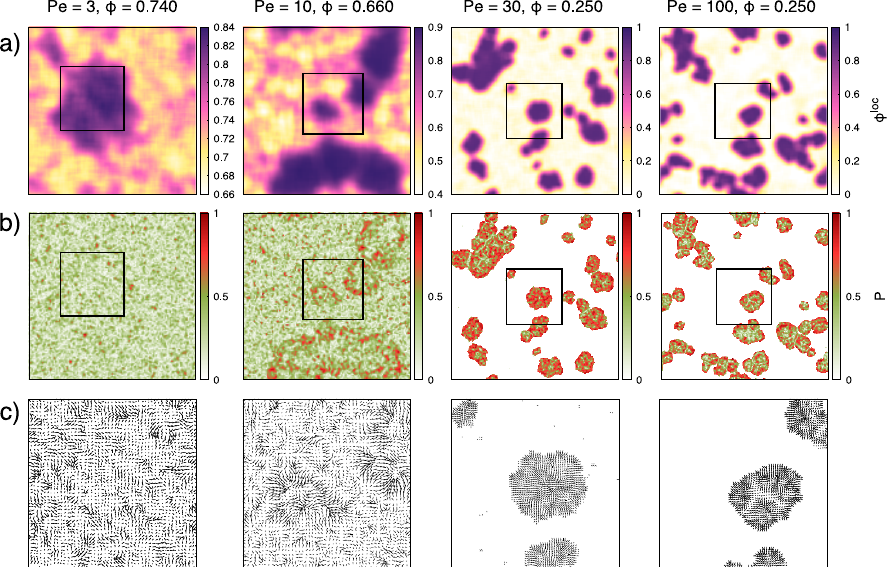}
  \caption{(a) Representative snapshots of the system for Pe $=3,10,30,100$, and global density corresponding to a dense phase occupying 25\% of the total surface. Density values are reported over each snapshot. Snapshots are colored according to the local surface fraction (see Fig.~\ref{fig:phase_diagram_confs}). (b) Same snapshots as in (a), colored according to the modulus of the local average polarization field $P$, obtained averaging the local polarization over coarse-graining length of $5\sigma_d$. (c) Polarization field for same conformations as (a-b), shown enlarged in a high density region of each snapshot. 
  This enlarged portion is indicated with a black square in the corresponding upper snapshots.
  For panels (a-b) the coloring code is reported at the right of each figure.
}
  \label{fig:pol_snap}
\end{figure}

We start by analyzing the polarization of the dense regions of dumbbells, by looking at the local average ${\mathbf P}$ of the self-propulsion direction ${\mathbf F}_a/F_a$, over a coarse-graining length of $5\sigma_d$, which we choose to enhance visualization. We analyze conformations in points of $\phi$-Pe space where the dense phase occupies $25\%$ of the total area (red curve in Fig.~\ref{fig:phase_diagram}) and more separate clusters appear. Snapshots of these conformations are shown in Fig.~\ref{fig:pol_snap}(a), colored according to the local density (panel (a)).

In panel Fig.~\ref{fig:pol_snap}(b), we show the same snapshots of Fig.~\ref{fig:pol_snap}(a) colored according to the modulus of the local average polarization $P$. Values near $1$ (red regions) imply a stronger local alignment of the active force directions. At Pe $=3$, a region of higher density (purple region) doesn't show any local alignment. At Pe $=10$, instead, regions at the cluster border become more polarized. At Pe $=30$ and $100$, these red regions of high polarization become even larger, and are again typically concentrated at the cluster border. These same observations can be appreciated by looking directly at the polarization field for an enlargement within the high density region of each snapshot, Fig.~\ref{fig:pol_snap}(c). Typically, for Pe $\gg1$ dumbbells point with their active force on average more towards the cluster center, with their polarization locked. Moreover, the polarization pattern can form a spiral, with the active force oriented at an angle with respect to the cluster's center. This breaking in the chiral symmetry can typically cause clusters to have a rotational motion~\cite{Suma14,Petrelli18}, as we will characterize in the next section.

\begin{figure}[t!]
  \centering
  \includegraphics[width=\textwidth]{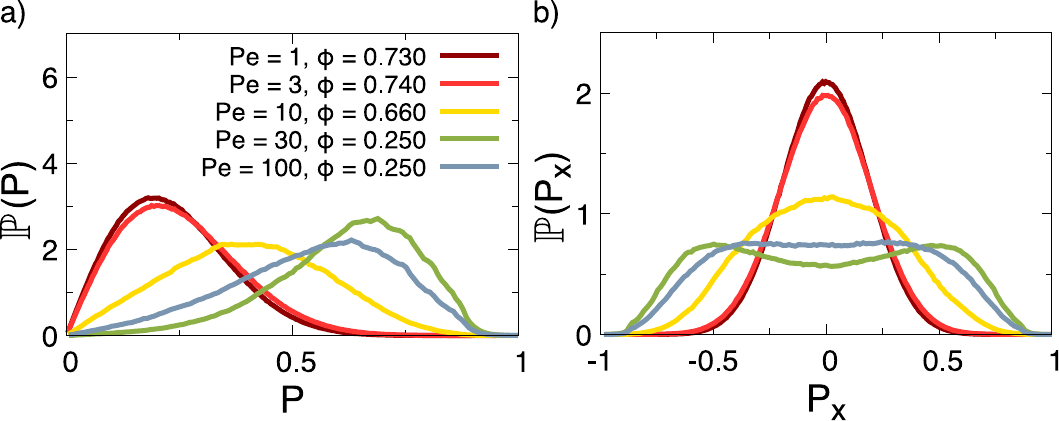}
  \caption{Probability distribution of (a) the modulus of the local polarization, $P$, and (b) its x-component, $P_x$, for points of the $\phi$-Pe space indicated in the key. These points are characterized by a dense phase occupying $25\%$ of the total surface (red line in Fig.~\ref{fig:phase_diagram} and snapshots in Fig.~\ref{fig:pol_snap}).
  }
  \label{fig:pol_pdfs}
\end{figure}

We now turn to a more through characterization of the polarization, by computing the probability distribution of the polarization $\mathbb{P}(P)$, Fig.~\ref{fig:pol_pdfs}(a). For small but not zero P\'eclet, in the regime Pe $\leq 3$, the distribution is compatible with a Maxwellian distribution, which means that no net local polarization is present. Increasing P\'eclet above Pe $=3$, we note a deviation from the Maxwellian behaviour, associated with the appearance of a peak at a non-null local polarization. This scenario is confirmed by the distribution of the x component of the local polarization, shown in Fig.~\ref{fig:pol_pdfs} (b). While it is normally distributed around zero for small P\'eclet, it develops fat tails at Pe $=10$ and non-zero, symmetric, components for Pe $=30,100$. These results are within the expectation that in the phase-separated region of the phase diagram, as activity becomes large enough, phase separation is mostly driven by persistence motion, rather than equilibrium-reminiscent liquid-hexatic coexistence, and dumbbells' interlocking, as described above, becomes relevant in the determination of clusters' internal structure, favouring local polarization.

Looking at Fig.~\ref{fig:pol_pdfs}, we note that the degree of polarization is slightly re-entrant for large P\'eclet: the peak of the probability distribution of $P$ (Fig.~\ref{fig:pol_pdfs} (a)) moves towards a higher polarization for Pe $\le 30$ and than back to a smaller polarization at Pe $=100$. This observation can be explained by looking at the structure of clusters formed at high P\'eclet in Fig.~\ref{fig:pol_snap}(c) for Pe $=30,100$, respectively. While the polarization pattern at Pe $=30$ develops around a single common center, typical clusters at higher Pe show a structure with patches of independent polarization, and non-polarized boundaries between them. This is due to a different mechanism of cluster formation: while at intermediate Pe clusters prevalently form through a process of nucleation and growth, at high Pe it becomes more likely that clusters have a stronger active motion, and thus can coalesce more, while preserving their quenched polarization structure, making more likely to form these types of multi-center polarization patterns.

%%%%%%%%%%%%%%%%%%%%%%%%%%%%%%%%%%%%%%%%%%
\subsection{Rotational properties of clusters}

As already observed for axially self-propelled dumbbells, clusters with a quenched polarization pattern are naturally subject to a persistent rotation motion~\cite{caporusso2024phase,Caporusso23}. Self-propulsion forces acting on each particle within the clusters contribute to a net force and torque on the center of mass of the cluster, which are generally non zero. Since we observed that the polarization pattern changes with Pe and presents non-trivial arrangements within the clusters, which is related to the clusters' size and shape, we studied the rotational motion as a function of the clusters' size, for different values of P\'eclet, again choosing densities where the dense phase occupies $25\%$ of the total area, and thus multiple clusters are available.

Clusters are identified by means of a DBSCAN algorithm~\cite{dbscan}.
The size of the clusters is measured as their radius of gyration $r_g$, computed as $N_c r_g^2 = \sum_{i=1}^{N_c} ({\mathbf r}_i-{\mathbf r}_{cm})^2$, where the index $i$ runs over the $N_c$ beads inside the cluster of mass $M_c=m_dN_c$ and position ${\mathbf r}_{cm}$. The angular velocity of the clusters is measured as $\boldsymbol{\omega}=\mathbf L/I$, with $I$ the moment of inertia with respect to the center of mass of the cluster and $\mathbf L=\sum_{i=1}^{N_c} ({\mathbf r}_i-{\mathbf r}_{cm}) \times m_d {\mathbf v}_i$ the modulus of the angular momentum. To measure the angular momentum, we use time-averaged particles velocities, over a time range $\Delta t=1$, which is large enough so that the random component of the instantaneous particles' velocity averages out, and smaller than the typical period of rotation of the clusters (the smallest period measured is of order $10^4$).
We perform this analysis for Pe $\ge 10$, where clusters have non-zero local polarization.

\begin{figure}[t!]
  \centering
  \includegraphics[width=\textwidth]{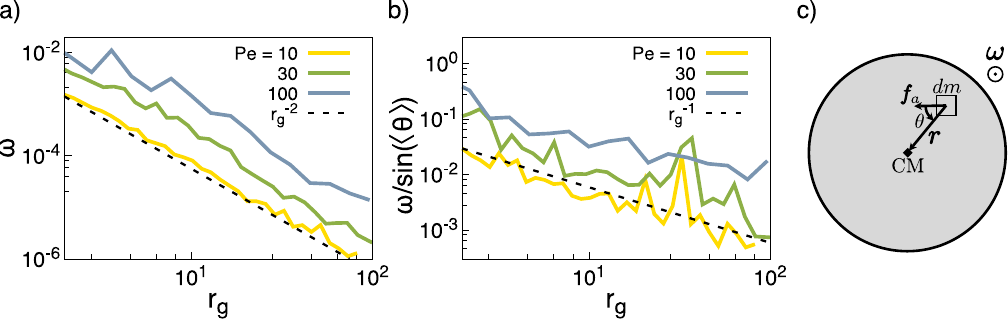}
  \caption{(a) Modulus of the rotating velocity of the clusters as a function of their radius of gyration, see the definition of these quantities in the text, for Pe $=10,30,100$. (b) Modulus of the rotating velocity of the clusters, normalized by the average polarization angle within the cluster, as defined in the text. (c) Graphical representation of a cluster (gray circle), highlighting the vector ${\mathbf r}$ connecting a mass element $dm$ to the cluster's center of mass, the active force $f_a$ acting on $dm$, and the angle $\theta$ between ${\mathbf f}_a$ and ${\mathbf r}$. With the assigned direction of ${\mathbf f}_a$, the cluster rotates counter-clockwise (see the direction of the vector $\boldsymbol{\omega}$ in the sketch).
  }
  \label{fig:omega}
\end{figure} 

Fig.~\ref{fig:omega} (a) displays the average of the modulus of the clusters' angular velocity, $\omega$, as a function of $r_g$ for Pe $=10,30,100$. We find here that $\omega \propto r_g^{-2}$, dashed line, for all Pe considered. This dependence is in stark contrast with the one found for axially propelled dumbbells, where instead $\omega \propto r_g^{-1}$~\cite{Suma14,Petrelli18}. In order to understand the origin of this dependence, we can develop a simplified continuum model, approximating a cluster as a two-dimensional rigid body with circular shape with constant density $\rho$, radius $R$ and area $A$, following a similar scheme as Ref.~\cite{caporusso2024phase}.

For each mass element $dm=\rho\, dA$ ($dA$ being the element area) on a circular cluster, the total force is $d\mathbf{f} = ({\mathbf{f}}_a - \gamma \mathbf{v}) dm$, comprised of the active force and the friction, with ${\mathbf{f}}_a$ the active force acting on the element, and $\mathbf{v}$ its velocity, both functions of the position $\mathbf{r}$ of the element. We take as origin of our coordinates the circle center of mass. We also assume  the cluster has a constant angular velocity around its center, $\boldsymbol{\omega}$, such that $\mathbf{v} = \boldsymbol{\omega} \times \mathbf{r}$ (with $\boldsymbol{\omega}$ perpendicular to the circle's plane). The total torque acting on the circle is obtained by integrating over the total area $A$ the torque $d\boldsymbol{\tau}=\mathbf{r} \times d\mathbf{f}$ acting on each mass element $dm$:
\begin{equation}
\begin{split}
    \boldsymbol{\tau} &= \int_A d\boldsymbol{\tau} = \int_A \mathbf{r} \times (\mathbf{\mathbf{f}}_a - \gamma \mathbf{v}) dm = \rho \int_A\, (\mathbf{r} \times \mathbf{\mathbf{f}}_a - \gamma\, \mathbf{r} \times (\boldsymbol{\omega} \times \mathbf{r})) \, dA\\
    &= \rho\int_A  \, \left(\mathbf{r} \times \mathbf{\mathbf{f}}_a - \gamma [(\mathbf{r} \cdot \mathbf{r}) \boldsymbol{\omega} - (\mathbf{r} \cdot  \boldsymbol{\omega})\mathbf{r}]\right) \, dA \\
    &= \rho\int_A  \, (\mathbf{r} \times \mathbf{\mathbf{f}}_a - \gamma r^2 \boldsymbol{\omega}) \, dA , 
\end{split}
\end{equation}
where in the first row we used the definition of $\mathbf{v}$, while in the second the triple product property of the cross product. Clearly $\boldsymbol{\tau}$ is parallel to $\boldsymbol{\omega}$ and perpendicular to the circle's plane. To simplify the formula, we now assume the active force per unit mass ${\mathbf{f}}_a$ has a constant magnitude $f_a$ throughout the disk, and maintains a constant angle $\theta$ relative to the position vector $\mathbf{r}$, see Fig.~\ref{fig:omega} (c) for a definition of $\theta$. Thus, the signed magnitude of the torque, evaluating the integral in polar coordinates $(r, \phi)$ becomes:
\begin{align}
    \tau &= \rho \int_0^{2\pi} d\phi \int_0^R dr \, (r^2 f_a \sin\theta-\gamma r^3 \omega)= \rho\pi\bigg (\frac{2}{3} f_a R^3 \sin\theta-\frac{1}{2}  \gamma \omega R^4\bigg ).
\end{align}
Finally, to achieve a steady-state rotation, the net torque must be zero~\cite{caporusso2024phase,Caporusso23}, providing a formula for the angular velocity:
\begin{equation} 
    \omega = \frac{4 f_a \sin\theta}{3 \gamma R}.
    \label{omega}
\end{equation}
As evident from Eq.~\ref{omega}, we can have multiple situations, based on how $\sin\theta$ scales with $R$. For our system of dumbbells with transverse propulsion, we need to have that the average orientation of the active force $\mathbf F_a$ with respect to the cluster center, $\langle\theta\rangle$, should be such that $\omega/\sin\langle\theta\rangle\sim 1/r_g$, in order to restore the correct dependence on $\omega$. 
Indeed, Fig.~\ref{fig:omega} (b) shows that this expected dependence is correctly found.

%%%%%%%%%%%%%%%%%%%%%%%%%%%%%%%%%%%%%%%%%%
\section{Conclusions}
\label{sec:conclusions}

We establish the phase diagram of a system of dumbbells, self-propelled with a constant force directed transverse to the axis of the molecule, using numerical methods. At zero self-propulsion, Pe $=0$, the system undergoes a liquid-hexatic phase transition of first-order, characterized by a coexistence region~\cite{Cugliandolo2017}. At Pe $>0$, we found that, similarly to axially self-propelled dumbbells, the coexistence region expands continuously, with the system showing phase separation between a dilute phase and a dense phase with hexatic ordering, at all P\'eclet numbers. Remarkably, for the case of transverse self-propulsion, the binodal region shows a widening towards lower densities, compared to axially self-propelled dumbbells. We motivate this behavior considering that dumbbells with a transverse force are i) able to form stable nuclei with less number of particles, and ii) more unlikely to diffuse away from the cluster's border. As a results, aggregation is highly enhanced for this model, compared to other dumbbell models.

Moreover, we show that for Pe $\geq 10$ the dense phase has polar order, with the self-propulsion direction in the interior of a cluster arranged into a spiralling pattern that breaks the chiral symmetry with respect to the cluster center of mass. The degree of internal polarization increases non-monotonically with Pe, since, for very large Pe, the internal polycrystalline structure disrupts dumbbells' alignment. This broken chiral symmetry drives the emergence of non zero active force and torque with respect to the center of mass of the clusters, leading to global translational and rotational motion of the clusters. In particular, we characterized the rotational motion, evaluating the dependence of the angular velocity, $\omega$, on the radius of gyration, $r_g$, of the clusters. We found that $\omega\sim r_g^{-2}$. This effective dependence can be explained by an analytical model which shows that, in general, $\omega\sim sin(\langle \theta \rangle)/r_g$, with $\theta$ the polarization angle with respect to the center of mass of the clusters.  For our system, $\langle \theta \rangle$ is inversely proportional to the clusters' radius of gyration, leaving overall to an effective $\omega \propto r_g^{-2}$ dependence.

\bibliography{dumbbellsbiblio_arxiv.bib}

\end{document}